# On transverse momenta and (pseudo)rapidity spectrum of the top quark in proton-proton collisions at 13 TeV

Li-Na Gao[*], Er-Qin Wang

Department of Physics, Taiyuan Normal University, Jinzhong, Shanxi 030619, China

**Abstract:** We study the transverse momenta and (pseudo)rapidity spectrum of the top quark and their decay products, the $t\bar{t}$ system, and the total number of jets in proton-proton (*pp*) collisions at 13 TeV by using the Tsallis-Pareto-type function and the three-source Landau hydrodynamic model, respectively. The related parameters, such as the effective temperature of the interacting system (T), the non-extensivity of the process (n) and the width ($\sigma$) of (pseudo)rapidity distribution are extracted.

## 1. Introduction

The top quark is an interesting particle. It was first found by the Tevatron detector of Fermi National Accelerator Laboratory in 1995[1,2]. As the heaviest particle in the Standard Model, the top quark is very different from the other quark. Its biggest mass is one of the hot spots for physicists. And, it is the only one quark which will decays before it become hadron. These unique properties make top quark occupies a very important position in particle physics.

The Large Hadron Collider (LHC) has created a lot of top quark events since 2010. Physicists could study the top quark in detail with the help of the LHC. By studying the final state particles produced in high energy collisions, physicists could obtain some information about the evolution of the collision system. The spectra of transverse momenta ($p_T$) and (pseudo)rapidity ($\eta / y$) are important quantities measured in experiments. Many models and functions are used to describe the transverse momenta and (pseudo)rapidity spectra.

In this paper, we use the Tsallis-Pareto-type function[3,4] and the three-source Landau hydrodynamic model[5] to describe the transverse momenta and (pseudo)rapidity spectra of the top quark, lepton and bottom quark produced in proton-proton (*pp*) collisions at the center of mass energy $\sqrt{s} = 13$ TeV measured at the parton level and particle level, in the full phase space and in the fiducial phase by CMS Collaboration at the LHC[6]. The values of related parameters are extracted and analyzed.

---

[*] E-mail : gao-lina@qq.com; gaoln@tynu.edu.cn

The present paper is organized as follows. We briefly introduce the Tsallis-Pareto-type function and the three-source Landau hydrodynamic model in section 2. The result of comparisons with experimental data is given in section 3. Moreover, we outlined the conclusions in section 4.

**2. The model and formalism**
2.1 The Tsallis-Pareto-type function

There are some models and functions could use to analyze the transverse momenta distribution, such as the Erlang distribution[7-9], the inverse power-law[10-13], the Schwinger mechanism[8,9,14,15], Lévy distribution[16], and the Tsallis statistics[17-22]. In a lot of models and formulas, the Tsallis-Pareto-type function is a good choice for transverse momenta spectra. Whether the relative high $p_T$ region or the relative low $p_T$ region, people could use the Tsallis-Pareto-type function to fit the experimental data well. So, in this work, we chose the Tsallis-Pareto-type function to describe the transverse momenta spectra. The Tsallis-Pareto-type function can be written as[4]:

$$\frac{d^2N}{dydp_T} = \frac{dN}{dy} Cp_T \left[1 + \frac{m_T - m_0}{nT}\right]^{-n} \quad (1)$$

Where $m_0$ is the rest mass of each particle, C and $m_T$ are given by:

$$C = \frac{(n-1)(n-2)}{nT[nT + (n-2)m_0]} \quad (2)$$

$$m_T = \sqrt{m_0^2 + p_T^2} \quad (3)$$

T and n are free parameters, T means the effective temperature of the interacting system and n denotes the non-extensivity of the process. We could obtain values of the two free parameters (T and n) by using the Tsallis-Pareto-type function to fit the $p_T$ spectra.

2.2 The three-source Landau hydrodynamic model

The three-source Landau hydrodynamic model has become a mature theoretical model in research the (pseudo)rapidity distributions at the high energy nuclear collisions. In our previous work, this model has described the experimental data successfully[5]. In present work, we will use it to fit the (pseudo)rapidity distributions of particles (top quark, lepton and bottom quark) produced in $pp$ collisions at $\sqrt{s} = 13$ TeV again. The following is a brief description of the three-source Landau hydrodynamic model.

The source means particles emission source. We think the rapidity distribution of particles produced in high energy collisions is contributed by the three emission sources. The three emission sources are a central source (C), a target source (T) and a projectile (P). The central source is located at the central of rapidity distribution and covers the whole rapidity range. The target source and projectile source are located at the left and right side of the central source, respectively. And, the target source and projectile source are revisions for the central source. For the rapidity distribution of particles produced in each emission sources, we can use a Gaussian form of the Landau solution to describe it[24-26]:

$$\frac{dN_{ch}}{dy} = \frac{N_0}{\sqrt{2\pi}\sigma_X} \exp\left(-\frac{(y-y_X)^2}{2\sigma_x^2}\right) \qquad (4)$$

Where $N_0$ is the normalization constant, y is rapidity, $\sigma$ denote the width of rapidity distribution, and X represents the type of emission source.

Actually, the three-source Landau hydrodynamic model can be written as a form of a superposition of three Gaussian forms of the Landau solution:

$$\frac{dN_{ch}}{dy} = \frac{N_0}{\sqrt{2\pi}} \left\{ \frac{k_T}{\sigma_T} \exp\left(-\frac{(y-y_T)^2}{2\sigma_T^2}\right) + \frac{k_C}{\sigma_C} \exp\left(-\frac{(y-y_C)^2}{2\sigma_C^2}\right) + \frac{k_P}{\sigma_P} \exp\left(-\frac{(y-y_P)^2}{2\sigma_P^2}\right) \right\} \qquad (5)$$

A given parameter $k$ is the contribution of each emission source, $k_T + k_C + k_P = 1$.

For the symmetric collision, we think $k_T = k_P$. Because of $y \approx \eta$ at very high energy, we could describe the pseudorapidity distributions of lepton by the following formula:

$$\frac{dN_{ch}}{d\eta} = \frac{N_0}{\sqrt{2\pi}} \left\{ \frac{k_T}{\sigma_T} \exp\left(-\frac{(\eta-\eta_T)^2}{2\sigma_T^2}\right) + \frac{k_C}{\sigma_C} \exp\left(-\frac{(\eta-\eta_C)^2}{2\sigma_C^2}\right) + \frac{k_P}{\sigma_P} \exp\left(-\frac{(\eta-\eta_P)^2}{2\sigma_P^2}\right) \right\} \qquad (6)$$

Significantly, as a revision of the central source, the contributions of the target source and projectile source are small.

**3. Comparisons with Experimental Data**

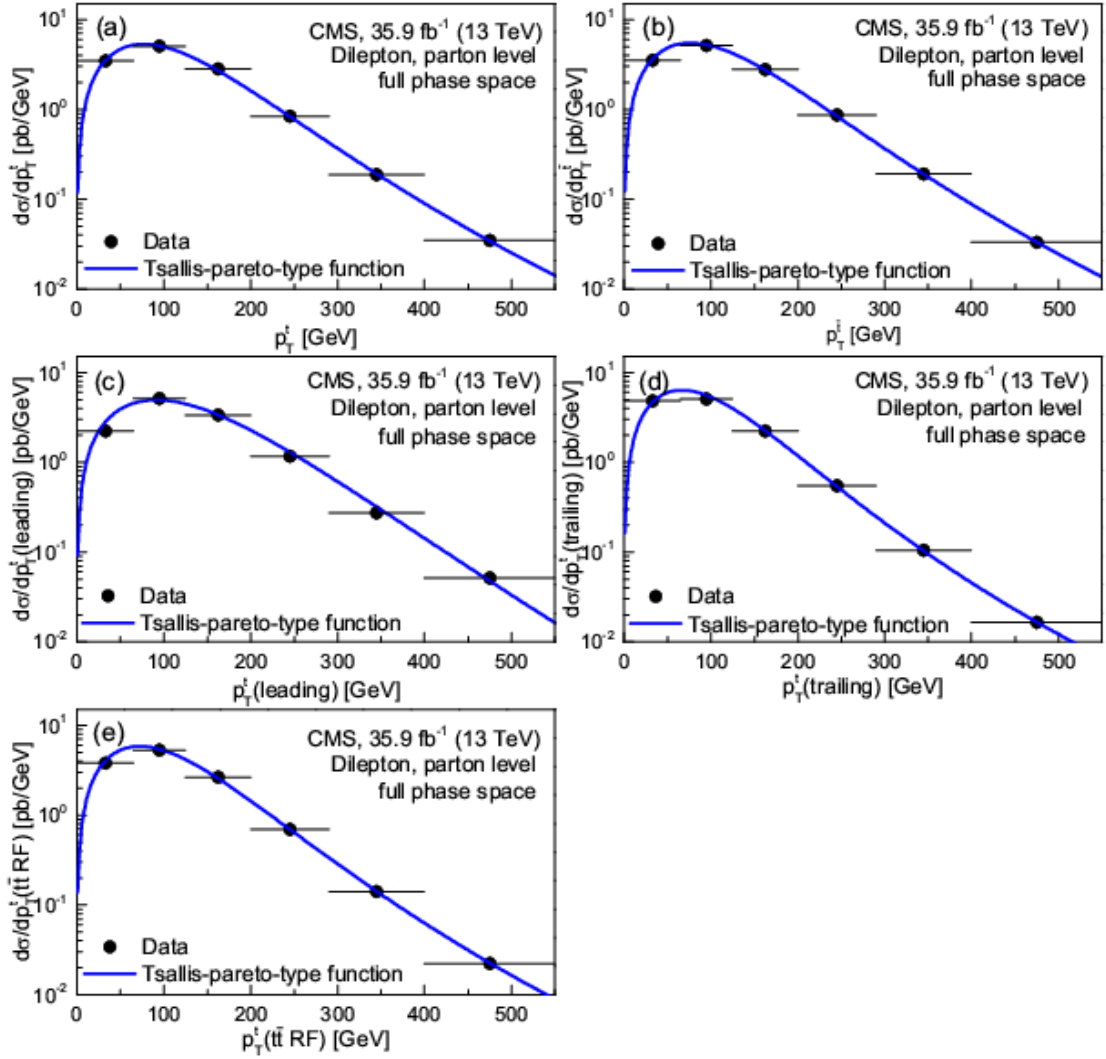

Figure 1: Transverse-momentum distribution of (a) the top quark ($p_T^t$) (b) the top antiquark ($p_T^{\bar{t}}$) (c) the top quark or top antiquark with largest $p_T$ ($p_T^t$(leading)) (d) the top quark or top antiquark with second-largest $p_T$ ($p_T^t$(trailing)) (e) the top quark in the rest frame of the $t\bar{t}$ system ($p_T^t(t\bar{t}$ RF)) at parton level in the full phase space produced in *pp* collisions at $\sqrt{s} = 13$ TeV. The solid circles represent the experimental data of the CMS Collaboration in literature[6], the curves are our results calculated by the Tsallis-pareto-type function.

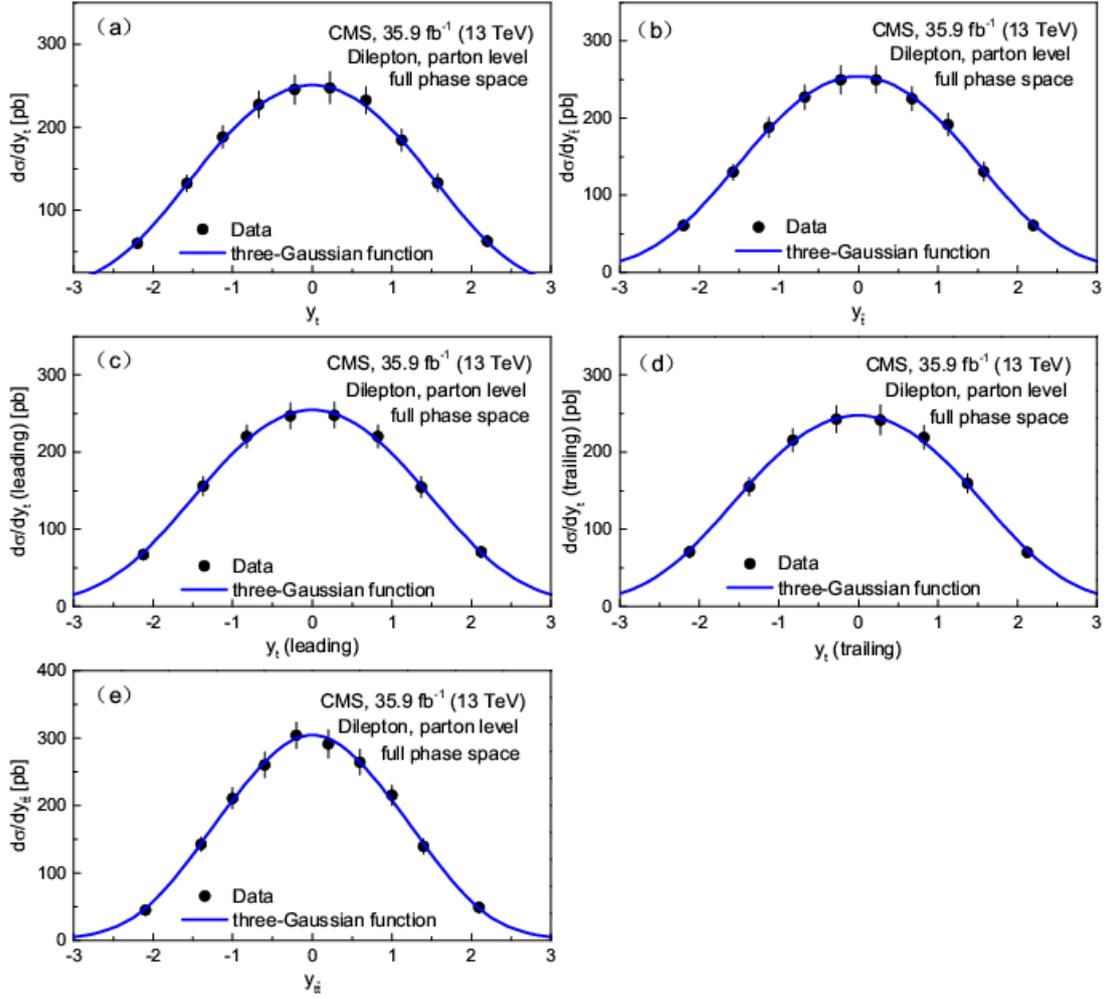

Figure 2: Rapidity distributions of (a) the top quark ($y_t$) (b) the top antiquark ($y_{\bar{t}}$) (c) the top quark or top antiquark with largest $p_T$ ($y_t$(leading)) (d) the top quark or top antiquark with second-largest $p_T$ ($y_t$(trailing)) (e) the $t\bar{t}$ system ($y_{t\bar{t}}$) at parton level in the full phase space produced in $pp$ collisions at $\sqrt{s}=13$ TeV. The solid circles represent the experimental data of the CMS Collaboration in literature[6], the curves are our results calculated by the three-Gaussian functions.

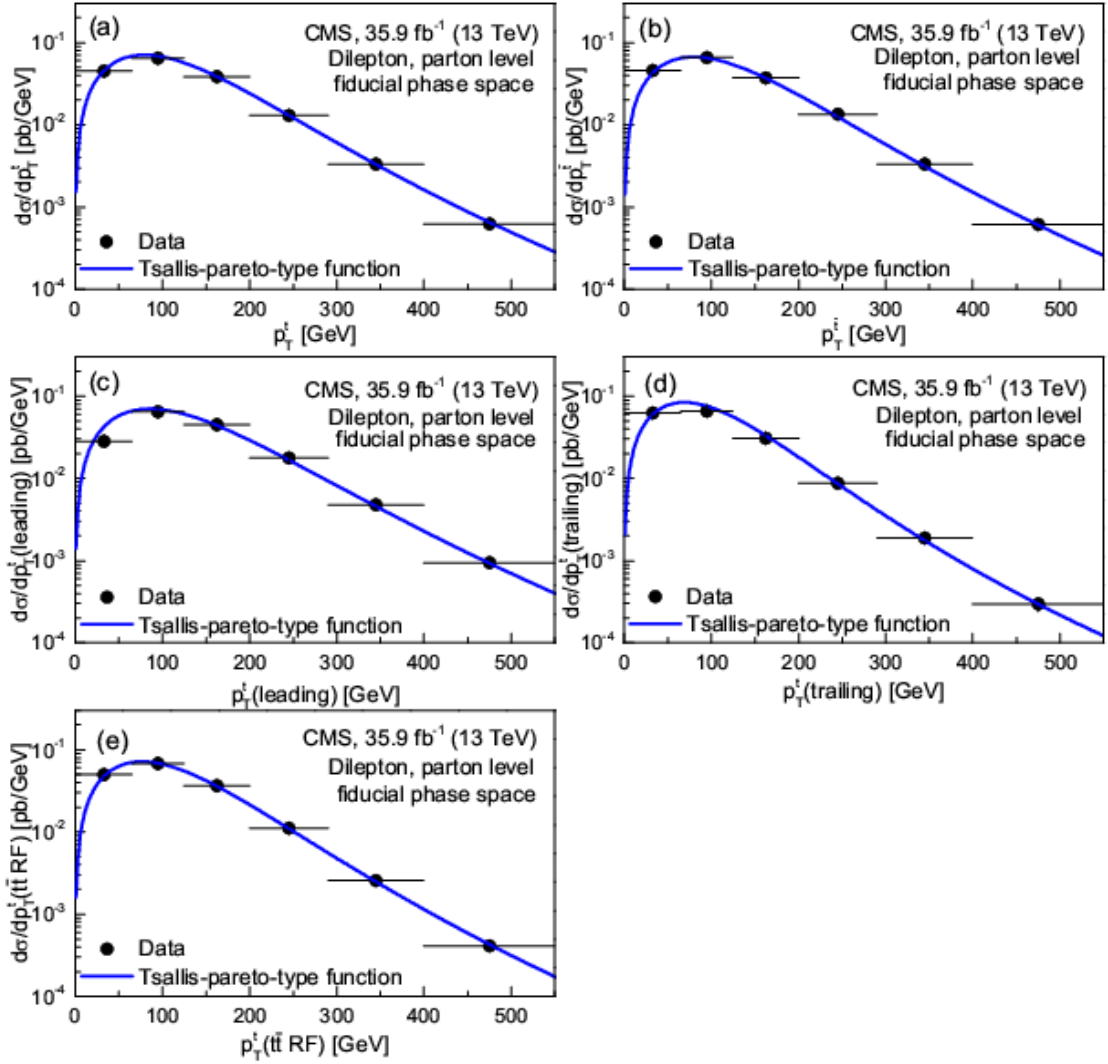

Figure 3: The same as Figure 1 but showing the results of transverse-momentum distribution of (a) $p_T^t$ (b) $p_T^{\bar{t}}$ (c) $p_T^t$ (leading) (d) $p_T^t$ (trailing) (e) $p_T^t$ ($t\bar{t}$ RF) at particle level in the fiducial phase space.

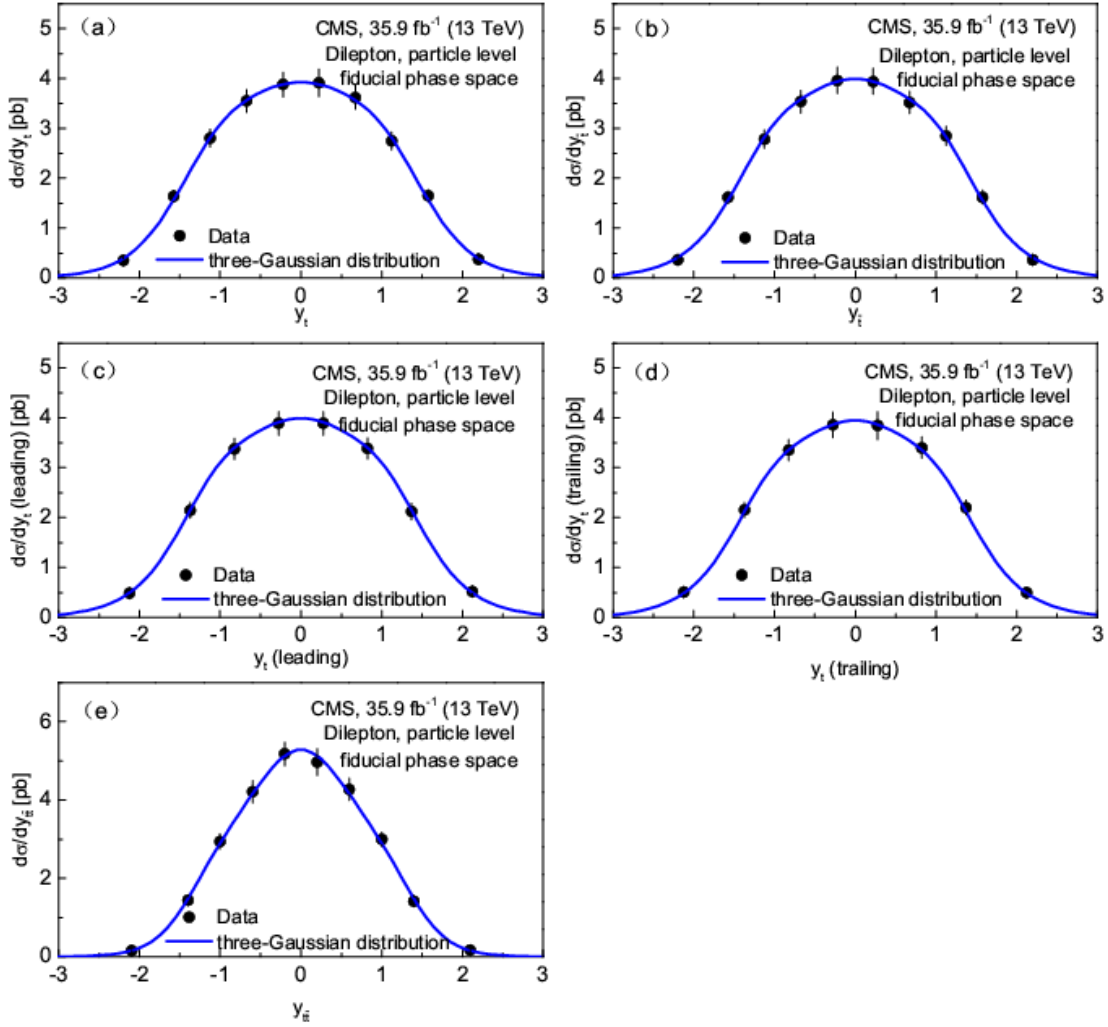

Figure 4: The same as Figure 2 but showing the results of (a) $y_t$ (b) $y_{\bar{t}}$ (c) $y_t$ (leading) (d) $y_t$ (trailing) (e) $y_{t\bar{t}}$ at particle level in the fiducial phase space.

Figure 1 and 2 show the transverse momenta and rapidity cross-sections of (a) the top quark ($t$) (b) the top antiquark ($\bar{t}$) (c) the top quark or top antiquark with largest $p_T$ ($t$ (leading)) (d) the top quark or top antiquark with second-largest $p_T$ ($t$ (trailing)) (e) the top quark in the rest frame of the $t\bar{t}$ system ($t\bar{t}$ RF) at parton level in the full phase space produced in $pp$ collisions at $\sqrt{s} = 13$ TeV. The same as Figure 1 and 2, but Figure 3 and 4 show the results at particle level in the fiducial phase space. In these four figures, the solid circles represent the experimental data recorded by the CMS experiment at the LHC and correspond to an integrated luminosity of 35.9 fb$^{-1}$ in 2016[6]. In figure 1 and 3, the curves are our results calculated by the Tsallis-pareto-type function. In figure 2 and 4, the curves show the calculate results of the three-Gaussian functions. Obviously, our results fit well with the experimental data. One can see that the Tsallis-Pareto-type function and the three-Gaussian functions are very helpful approaches to describe the transverse momenta and rapidity cross-sections of particles, respectively. The related parameters values (n, T, k, $y_P$, $\sigma$) extracted from the transverse momenta and rapidity cross-sections and $\chi^2/dof$ are given in table 1 and 2. We could found the values of n and T of $t$(leading) is the biggest in the five type particles at parton and particle level in the full and fiducial phase space. The values of $y_P$ are the same in the scope of consideration. The values of k, $\sigma(C)$ and $\sigma(P)$ are basically the same within the margin of error in the $t$, $\bar{t}$, $t$ (leading) and $t$ (trailing), but this three parameters values decreased significantly in the rest frame of the $t\bar{t}$ system.

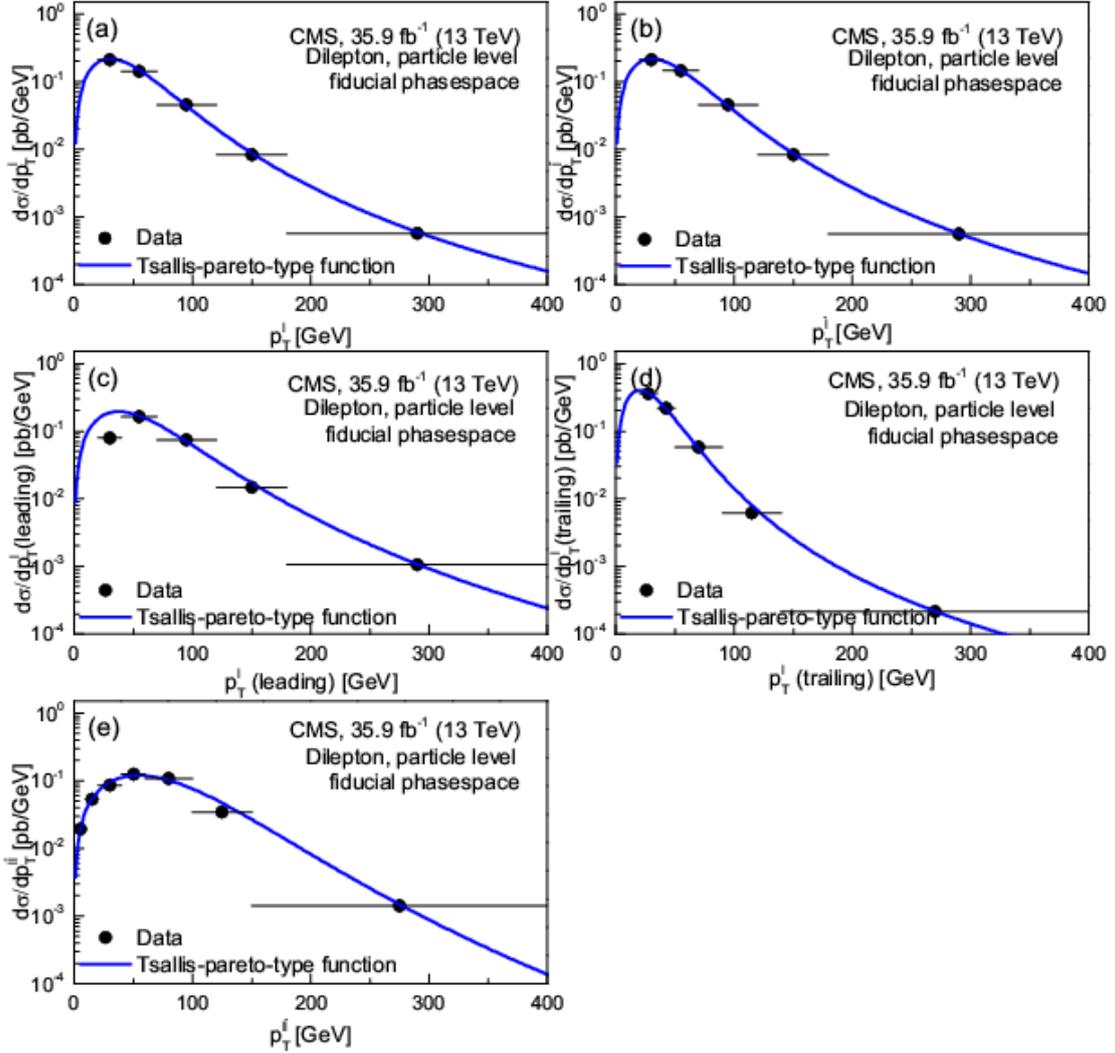

Figure 5: The same as Figure 1 but showing the results of transverse-momentum distribution of (a) the lepton ($p_T^l$) (b) the antilepton ($p_T^{\bar{l}}$) (c) the lepton or antilepton

with largest $p_T$ ($p_T^l$ (leading)) (d) the lepton or antilepton with second-largest $p_T$ ($p_T^l$ (trailing)) (e) the dilepton system ($p_T^{l\bar{l}}$) at particle level in the fiducial phase space.

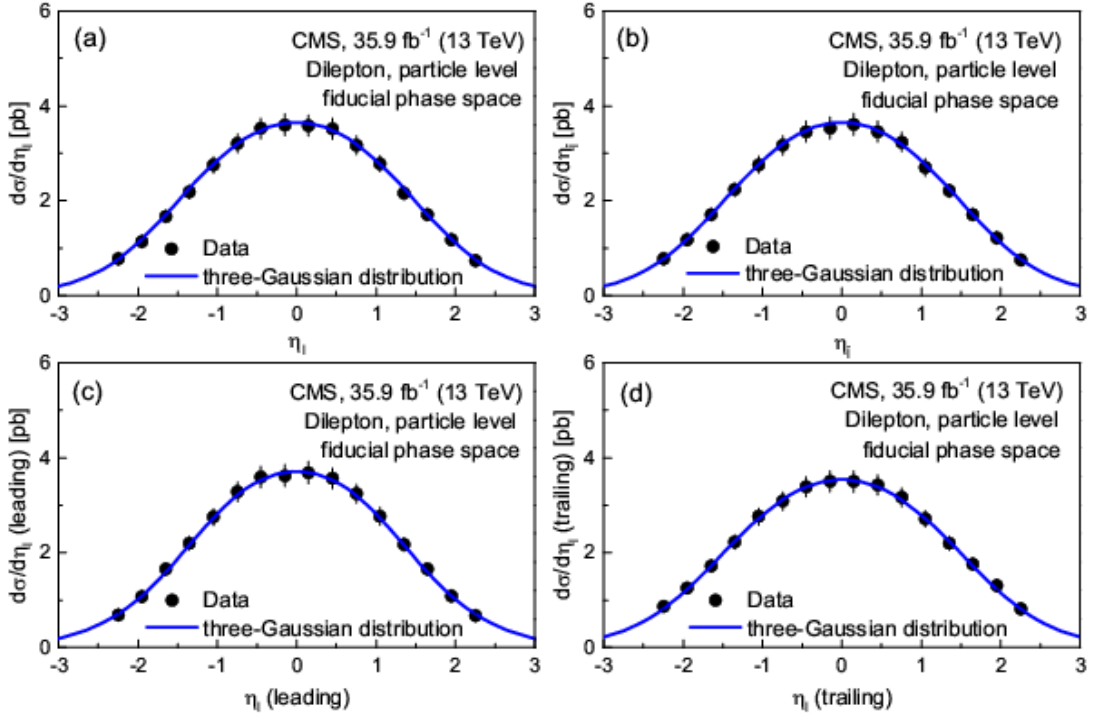

Figure 6: The same as Figure 2 but showing the results of pseudorapidity distribution of (a) the lepton ($\eta_l$) (b) the antilepton ($\eta_{\bar{l}}$) (c) the lepton or antilepton with largest $p_T$ ($\eta_l$ (leading)) (d) the lepton or antilepton with second-largest $p_T$ ($\eta_l$ (trailing))

Figure 5 and 6 show the transverse momenta and rapidity cross-sections of (a) the lepton ($l$), (b) the antilepton ($\bar{l}$), (c) the lepton or antilepton with largest $p_T$ ($l$ (leading)), (d) the lepton or antilepton with second-largest $p_T$ ($l$ (trailing)), and Figure 5(e) shows the transverse momenta cross-sections of the dilepton system ($l\bar{l}$) at particle level in the fiducial phase space produced in *pp* collisions at $\sqrt{s} = 13$ TeV. The solid circles and curves represent the experimental data and our calculate results, respectively. One can see that the results calculated by using the theoretical functions are in agreement with the experimental data. The related parameters are extracted and listed in the Table 1(5(a)-(e)) and 2 (6(a)-(d)). The values of n and T of $l\bar{l}$ system is bigger than the other particles. The values of $k$, $y_P$, $\sigma(C)$ and $\sigma(P)$ of the considered particles are not different obviously.

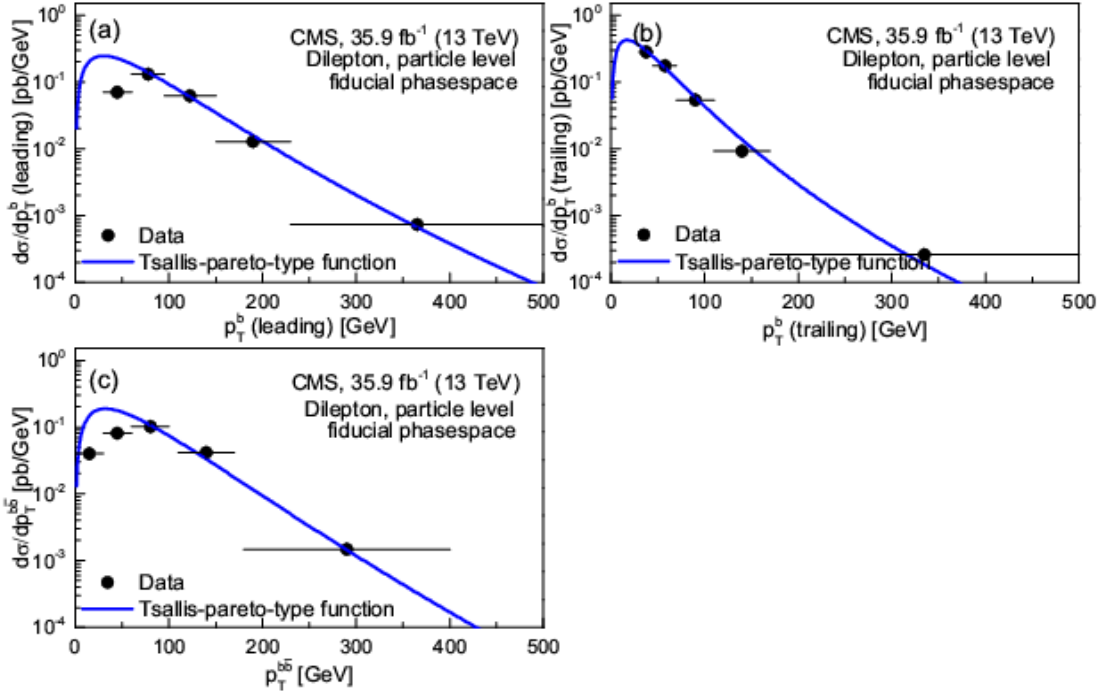

Figure 7: The same as Figure 1 but showing the results of transverse-momentum distribution of (a) the b jet with largest $p_T$ ($p_T^b$ (leading)) (b) the b jet with second-largest $p_T$ ($p_T^b$ (trailing)) (c) the $b\bar{b}$ system ($p_T^{b\bar{b}}$) at particle level in the fiducial phase space.

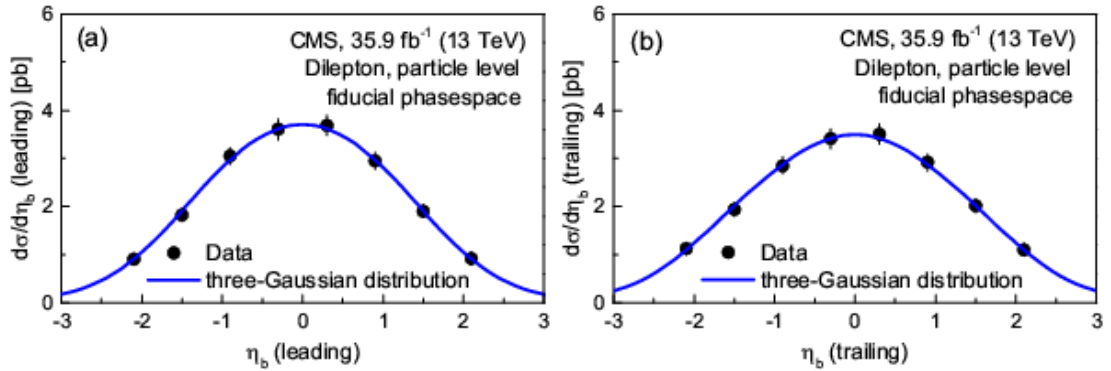

Figure 8: The same as Figure 2 but showing the results of pseudorapidity distribution of (a) the b jet with largest $p_T$ ($\eta_b$ (leading)) (d) the b jet with second-largest $p_T$ ($\eta_b$ (trailing))

The results of transverse-momentum and rapidity cross-sections of (a) the $b$ jet with largest $p_T$ ($b$ (leading)) (b) the $b$ jet with second-largest $p_T$ ($b$ (trailing)) are showed in Figure 7 and 8, and the Figure 7(c) shows the transverse-momentum cross-sections of the $b\bar{b}$ system at particle level in the fiducial phase space produced in $pp$ collisions at $\sqrt{s} = 13$ TeV. We have extracted the related parameters by describe the transverse-momentum and rapidity cross-sections distributions, and listed them in Table 1 (7(a)-(c)) and 2 (8(a)-(b)). The values of parameters are not an obvious relationship.

Generally, the initial temperature ($T_i$) and mean transverse momentum ($\langle p_T \rangle$) are important physical quantities to understand the excitation degree of the interacting system. The two physical quantities are not depend on the selected models and functions, they only dependent on the experimental data. According to the literatures [27-29], $T_i$ can be described by the ratio of root-mean-square $p_T$ to $\sqrt{2}$ ($\sqrt{\langle p_T^2 \rangle/2}$) approximately. So, we calculated the values of $\langle p_T \rangle$ and $\sqrt{\langle p_T^2 \rangle/2}$ by the curves in Figure 1, 3, 5, 7, and listed them in the Table 3. Figure 9 and 10 show the $\langle p_T \rangle$ and $\sqrt{\langle p_T^2 \rangle/2}$ of each particle. From Figure 9 and 10, we could found that the excitation degree of the interacting system produced the $t\bar{t}$ system is higher than other system.

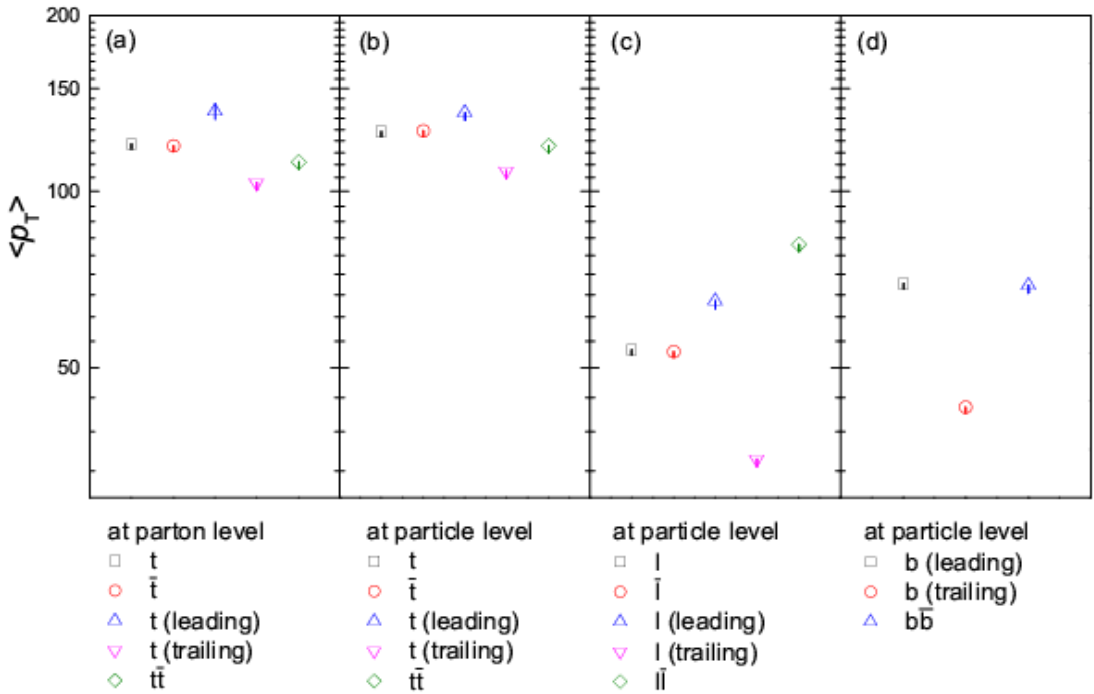

Figure 9: The values of $\langle p_T \rangle$ for different particles.

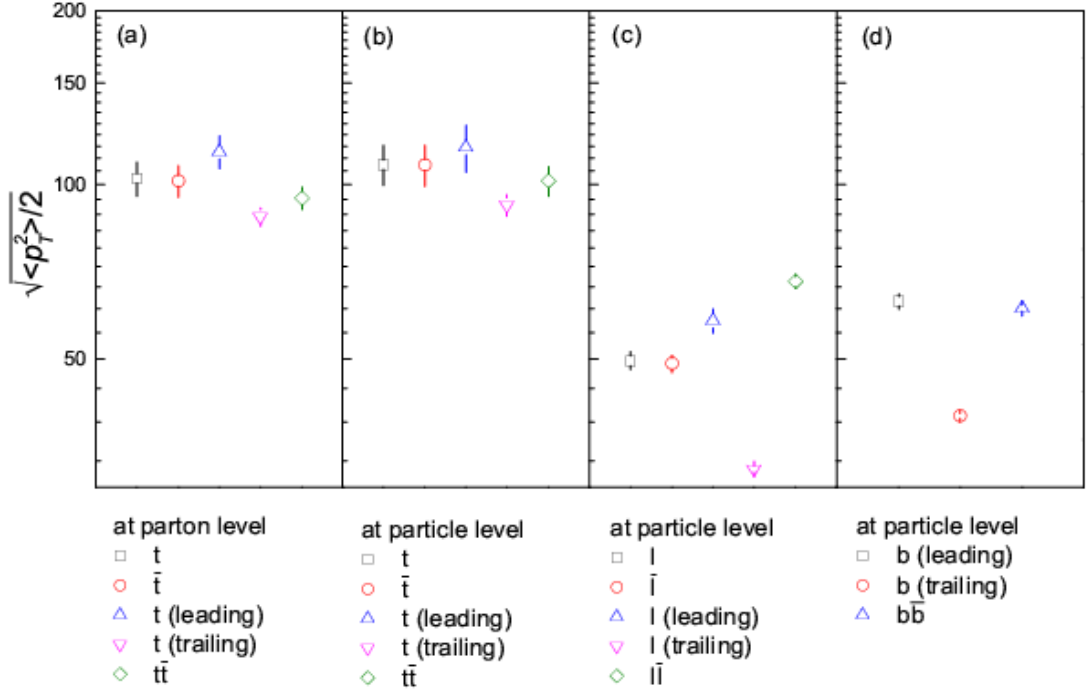

Figure 10: The same as figure 9 but showing the values of $\sqrt{\langle p_T^2 \rangle / 2}$.

In order to find the relationship between the free parameter and collision energy, we have extracted the related parameters from the transverse momenta and rapidity cross-sections at 7 TeV and 8 TeV (Figure 11-16), together. In Figure 11 and 12, the solid circles represent the experimental data recorded by the CMS experiment at 7 TeV and correspond to an integrated luminosity of 5.0 fb$^{-1}$ in 2011[30]. In Figure 13-16, the solid circles represent the experimental data recorded by the CMS experiment at 8 TeV and correspond to an integrated luminosity of 19.7 fb$^{-1}$ in 2012[31]. From the transverse momenta and rapidity cross-sections of final state particles, we extracted the values of free parameters and listed them in Table 1 and 2. And, we plot the T, n and $\sigma$ values listed in Table 1 and 2 in Figure 17. In the panel, the symbols represent the values of T, n and $\sigma$, and the lines are linear fitting functions. The intercepts, slopes and $\chi^2/dof$ corresponding to the lines are listed in Table 4. From the Figure 17 and Table 4, we found the values of T shows a slight increased with the collision energy increased. $\sigma(C)$ shows increased with the collision energy increased. The values of n about *t* and *l* do not show obvious change and the values of n about *b* decreased when the collision energy increased.

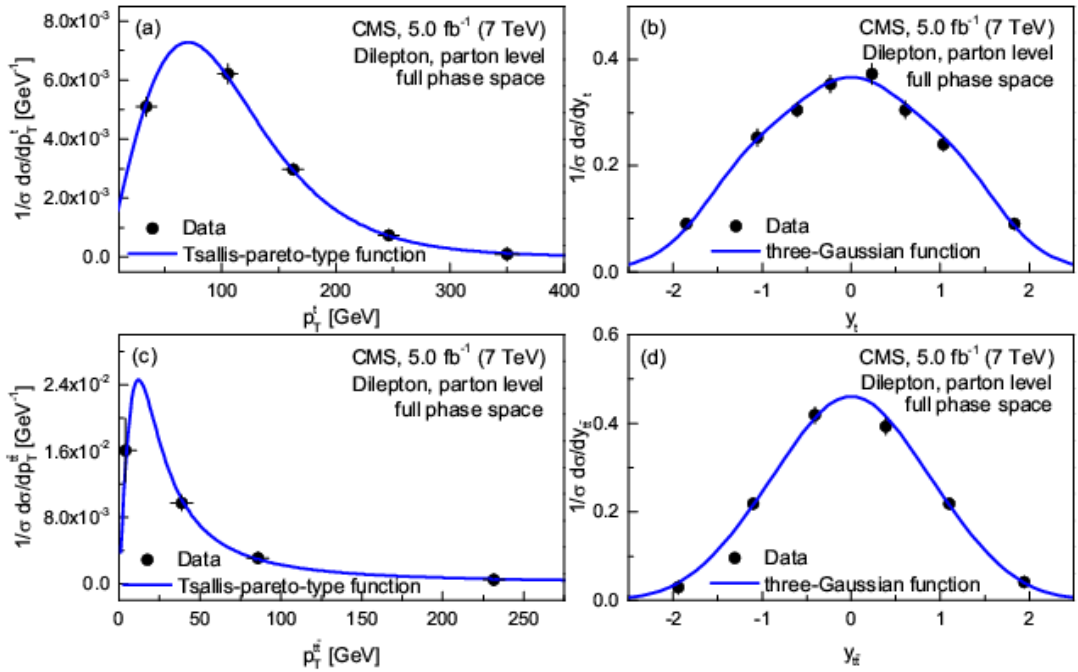

Figure 11: The same as Figure 1 and 2, but showing the results of $t$ and $t\bar{t}$ produced in $pp$ collisions at parton level in the full phase space at $\sqrt{s} = 7$ TeV.

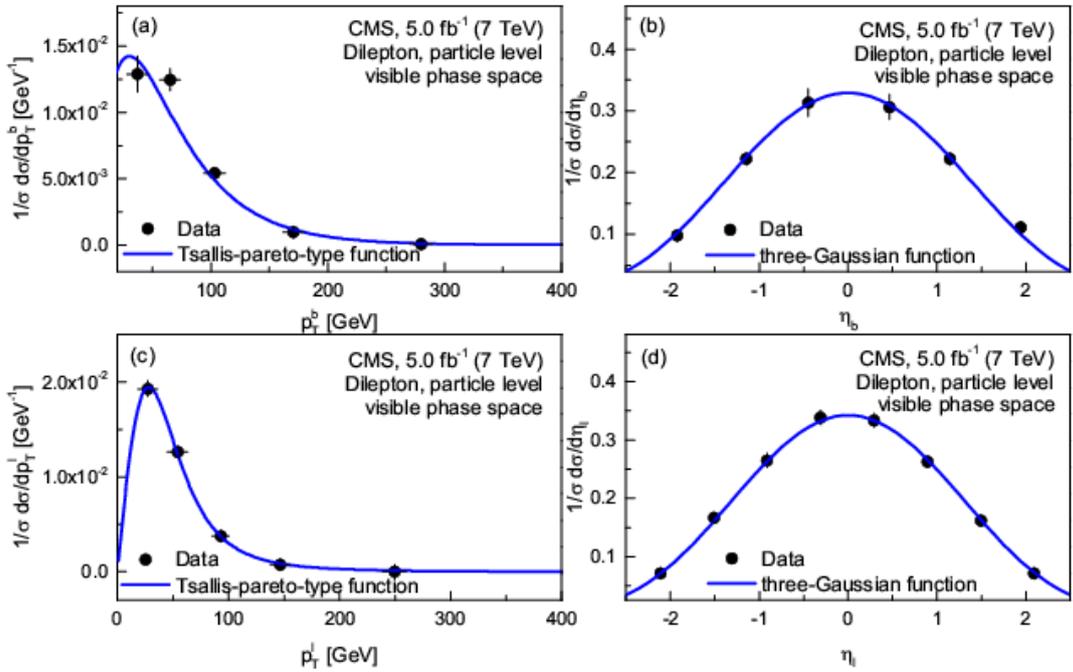

Figure 12: The same as Figure 1 and 5, but showing the results of $b$ and $l$ produced in $pp$ collisions at particle level in the visible phase space at $\sqrt{s} = 7$ TeV.

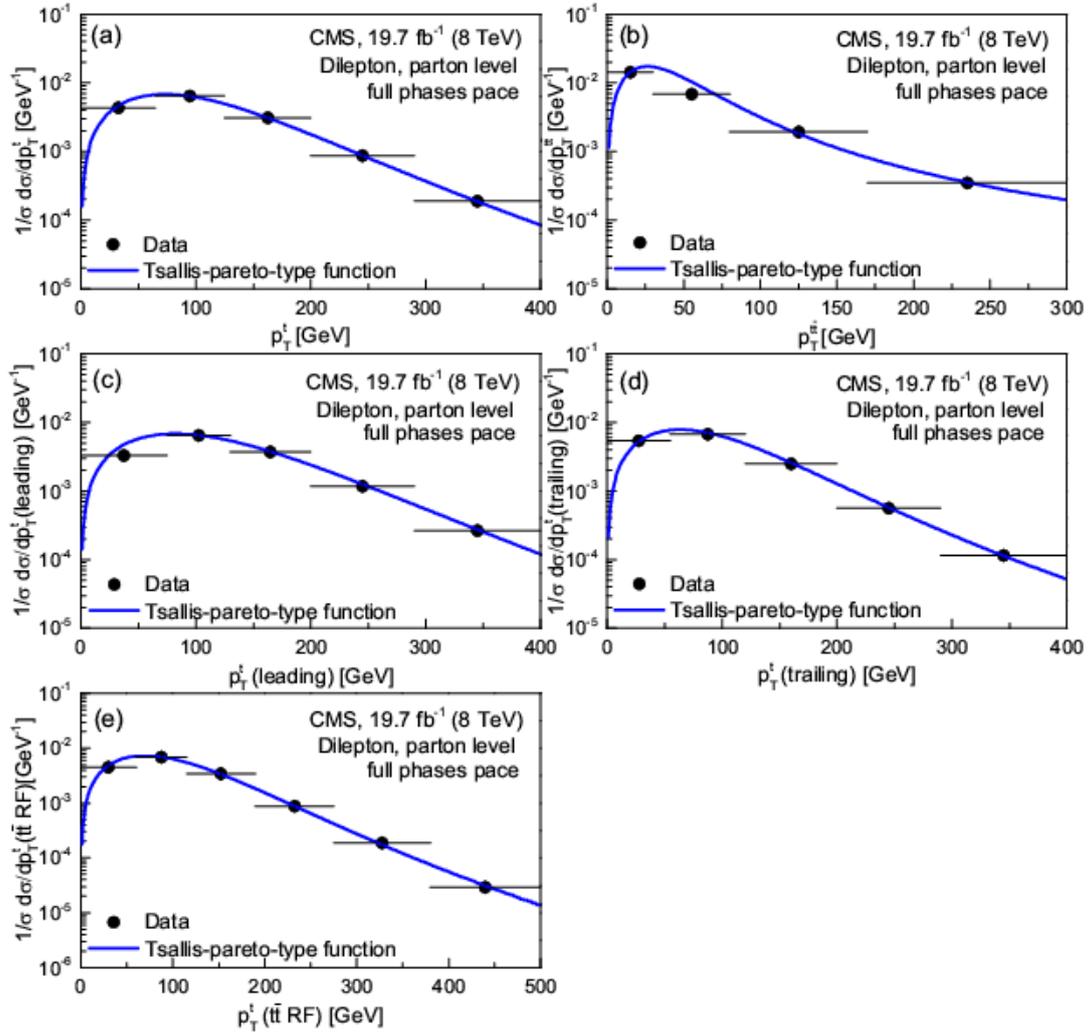

Figure 13: The same as Figure 1 but showing the results of transverse-momentum distribution of (a) $p_T^t$ (b) $p_T^{t\bar{t}}$ (c) $p_T^t$ (leading) (d) $p_T^t$ (trailing) (e) $p_T^t$ ($t\bar{t}$ RF) at parton level in the full phase space at $\sqrt{s}=8$ TeV.

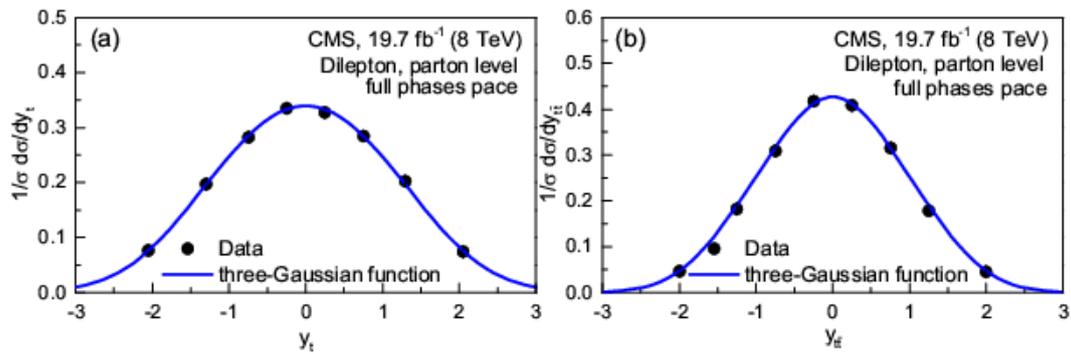

Figure 14: The same as Figure 2 but showing the results of rapidity distribution of $t$ and $t\bar{t}$ at parton level in the full phase space at $\sqrt{s}=8$ TeV.

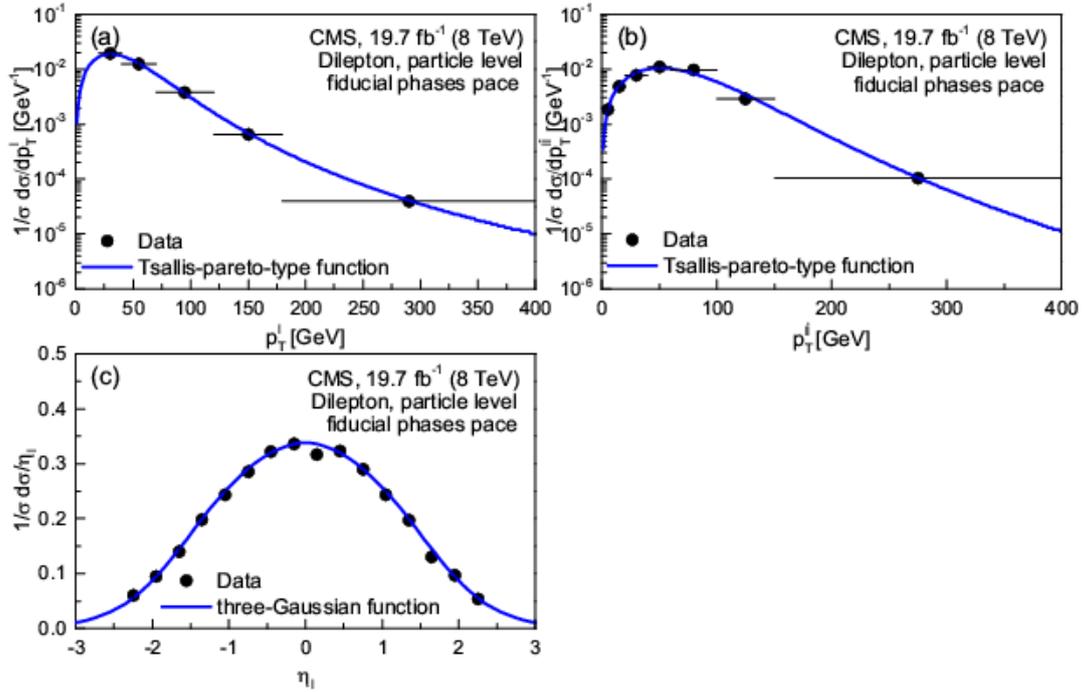

figure 15: The same as Figure 1and 5, but showing the results of $l$ and $l\bar{l}$ produced in *pp* collisions at particle level in the fiducial phase space at $\sqrt{s}=8$ TeV.

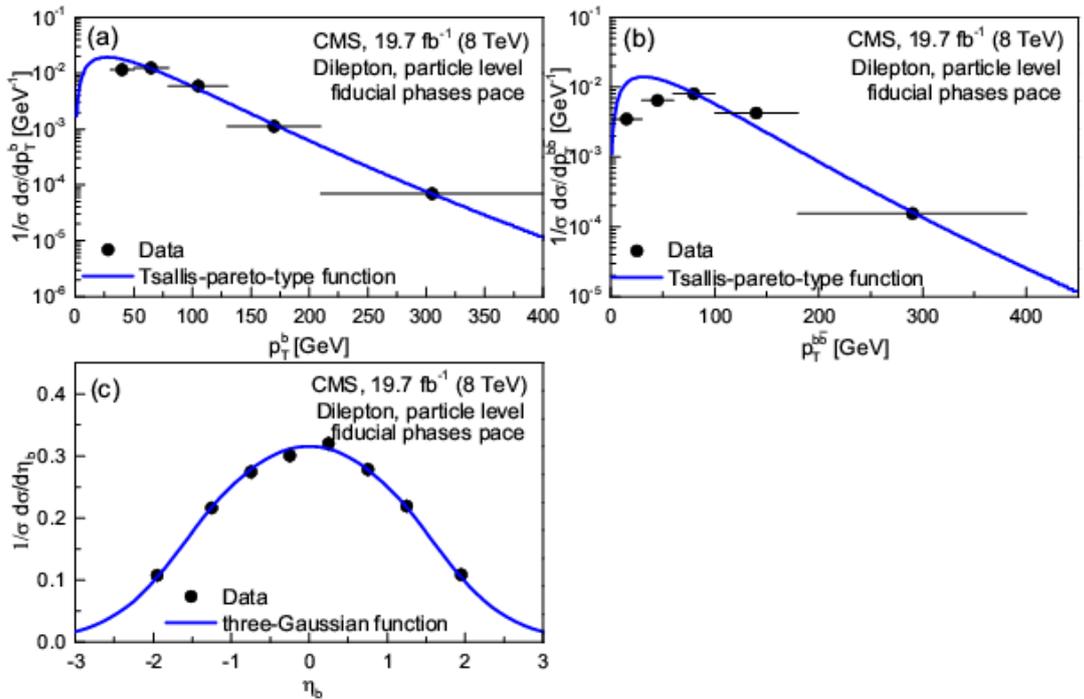

Figure 16: The same as Figure 1and 5, but showing the results of $b$ and $b\bar{b}$ produced

in *pp* collisions at particle level in the fiducial phase space at $\sqrt{s} = 8$ TeV.

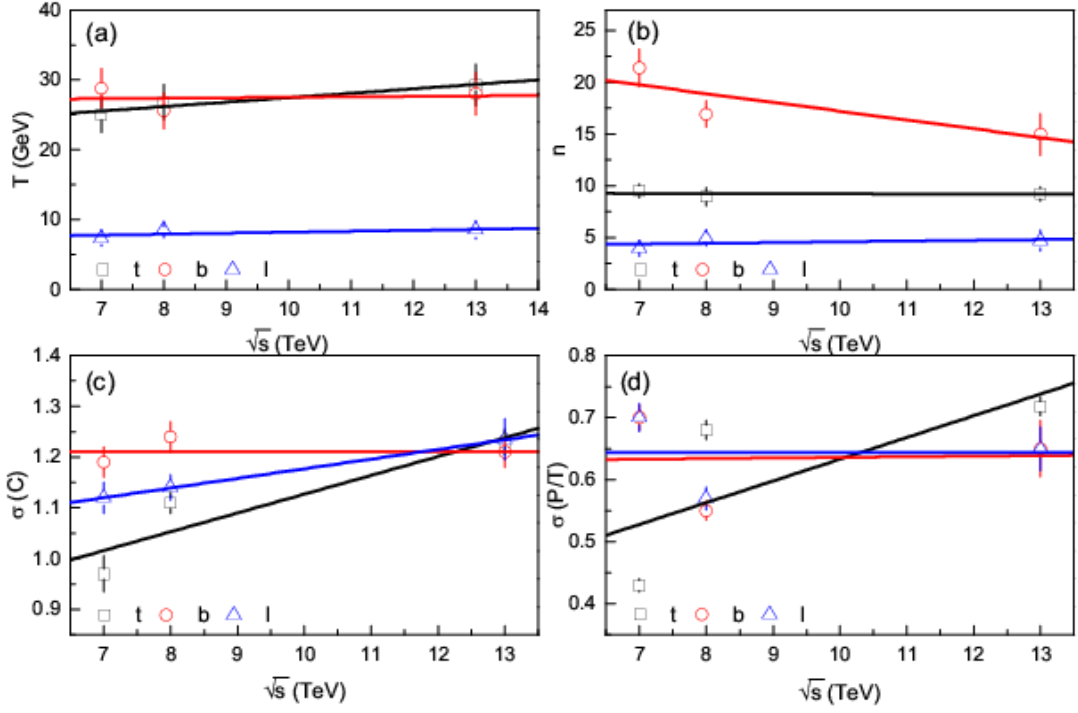

Figure 17: The relationship of free parameter and collision energy.

Table 1: Values of parameters and $\chi^2/dof$ corresponding to the curves in Figure 1, 3, 5, 7, 11, 12, 13, 15 and 16.

| Figure | Type | n | T (GeV) | $\chi^2/dof$ |
|---|---|---|---|---|
| | | 13 TeV | | |
| 1 (a) | $t$ | 9.23±0.65 | 29.26±2.98 | 0.021 |
| 1 (b) | $\bar{t}$ | 9.32±0.68 | 29.02±2.58 | 0.012 |
| 1 (c) | $t$ (leading) | 22.00±0.93 | 41.56±3.54 | 0.411 |
| 1 (d) | $t$ (trailing) | 8.05±0.53 | 22.25±2.13 | 0.037 |
| 1 (e) | $t$ ($t\bar{t}$ RF) | 9.22±0.62 | 26.25±2.25 | 0.022 |
| 3 (a) | $t$ | 8.80±0.57 | 31.00±2.85 | 0.018 |
| 3 (b) | $\bar{t}$ | 9.20±0.65 | 31.50±3.00 | 0.083 |
| 3 (c) | $t$ (leading) | 10.20±0.70 | 36.00±2.95 | 0.732 |
| 3 (d) | $t$ (trailing) | 8.20±0.52 | 24.00±2.20 | 0.150 |
| 3 (e) | $t$ ($t\bar{t}$ RF) | 9.50±0.67 | 29.20±2.50 | 0.030 |
| 5 (a) | $l$ | 4.72±1.00 | 8.50±1.30 | 0.056 |
| 5 (b) | $\bar{l}$ | 4.80±1.10 | 8.55±1.30 | 0.036 |
| 5 (c) | $l$ (leading) | 5.80±1.25 | 13.00±2.60 | 3.517 |
| 5 (d) | $l$ (trailing) | 4.15±0.85 | 3.90±0.70 | 0.324 |
| 5 (e) | $l\bar{l}$ | 7.88±1.55 | 14.50±3.00 | 0.611 |
| 7 (a) | b (leading) | 15.00±2.00 | 28.00±3.00 | 6.550 |
| 7 (b) | b (trailing) | 10.00±1.30 | 15.20±2.30 | 1.101 |
| 7 (c) | $b\bar{b}$ | 23.00±2.80 | 30.00±3.00 | 13.903 |
| | 7 TeV | | | |
| 11 (a) | $t$ | 9.52±0.68 | 25.00±2.50 | 0.024 |
| 11 (c) | $\bar{t}$ | 1.40±0.15 | 0.27±0.07 | 0.078 |
| 12 (a) | $b$ | 21.40±1.80 | 28.80±2.80 | 0.406 |
| 12 (c) | $l$ | 3.95±0.40 | 7.30±0.90 | 0.104 |
| | 8 TeV | | | |
| 13 (a) | $t$ | 8.96±0.86 | 26.82±2.52 | 0.064 |
| 13 (b) | $t\bar{t}$ | 2.32±0.30 | 3.10±0.40 | 1.206 |
| 13 (c) | $t$ (leading) | 13.54±1.16 | 33.68±2.80 | 1.144 |
| 13 (d) | $t$ (trailing) | 7.12±0.82 | 20.23±2.10 | 0.061 |
| 13 (e) | $t$ ($t\bar{t}$ RF) | 9.35±0.90 | 24.45±2.50 | 0.116 |
| 15 (a) | $l$ | 4.97±0.57 | 8.50±0.50 | 59.839 |
| 15 (b) | $l\bar{l}$ | 6.80±0.80 | 12.60±1.10 | 1.144 |
| 16 (a) | $b$ | 16.95±1.25 | 25.58±2.53 | 0.988 |
| 16 (b) | $b\bar{b}$ | 16.32±1.08 | 29.06±2.86 | 19.317 |

Table 2: Values of paramrters and $\chi^2/dof$ corresponding to the curves in Figure 2, 4, 6, 8, 11, 12, 14, 15, 16.

| Figure | Type | $k_C$ | $y_P/\eta_P$ | $\sigma(C)$ | $\sigma(P)$ | $\chi^2/dof$ |
|---|---|---|---|---|---|---|
| | | | 13 TeV | | | |
| 2 (a) | $t$ | 0.887±0.017 | 1.354±0.074 | 1.230±0.025 | 0.718±0.015 | 0.013 |
| 2 (b) | $\bar{t}$ | 0.892±0.020 | 1.354±0.074 | 1.230±0.025 | 0.718±0.018 | 0.026 |
| 2 (c) | $t$ (leading) | 0.908±0.022 | 1.354±0.074 | 1.245±0.030 | 0.725±0.018 | 0.019 |
| 2 (d) | $t$ (trailing) | 0.892±0.018 | 1.354±0.074 | 1.265±0.033 | 0.720±0.015 | 0.012 |
| 2 (e) | $t\bar{t}$ | 0.935±0.025 | 1.354±0.074 | 1.040±0.025 | 0.518±0.008 | 0.020 |
| 4 (a) | $t$ | 0.840±0.020 | 1.100±0.060 | 1.000±0.020 | 0.470±0.006 | 0.037 |
| 4 (b) | $\bar{t}$ | 0.875±0.022 | 1.100±0.060 | 1.020±0.020 | 0.425±0.008 | 0.102 |
| 4 (c) | $t$ (leading) | 0.875±0.022 | 1.100±0.060 | 1.020±0.022 | 0.425±0.007 | 0.053 |
| 4 (d) | $t$ (trailing) | 0.875±0.025 | 1.100±0.060 | 1.020±0.020 | 0.425±0.008 | 0.032 |
| 4 (e) | $t\bar{t}$ | 0.905±0.027 | 1.100±0.060 | 0.780±0.018 | 0.355±0.005 | 0.232 |
| 6 (a) | $l$ | 0.915±0.103 | 1.225±0.300 | 1.235±0.040 | 0.650±0.035 | 0.027 |
| 6 (b) | $\bar{l}$ | 0.912±0.100 | 1.225±0.300 | 1.235±0.040 | 0.672±0.040 | 0.014 |
| 6 (c) | $l$ (leading) | 0.950±0.110 | 0.990±0.285 | 1.235±0.030 | 0.560±0.030 | 0.033 |
| 6 (d) | $l$ (trailing) | 0.918±0.105 | 1.265±0.310 | 1.268±0.035 | 0.670±0.040 | 0.040 |
| 8 (a) | $b$ (leading) | 0.930±0.110 | 1.225±0.280 | 1.210±0.050 | 0.650±0.045 | 0.012 |
| 8 (b) | $b$ (trailing) | 0.915±0.105 | 1.550±0.300 | 1.270±0.050 | 0.680±0.060 | 0.012 |
| | | | 7 TeV | | | |
| 11 (b) | $t$ | 0.880±0.110 | 1.290±0.328 | 0.970±0.015 | 0.430±0.010 | 0.088 |
| 11 (d) | $t\bar{t}$ | 0.950±0.130 | 1.290±0.315 | 0.840±0.010 | 0.450±0.013 | 0.056 |
| 12 (b) | $b$ | 0.920±0.125 | 1.290±0.325 | 1.190±0.030 | 0.700±0.020 | 0.040 |
| 12 (c) | $l$ | 0.904±0.110 | 1.290±0.320 | 1.120±0.030 | 0.700±0.022 | 0.009 |
| 8 TeV | | | | | | |
| 14 (a) | $t$ | 0.900±0.080 | 1.290±0.290 | 1.110±0.020 | 0.680±0.015 | 0.028 |
| 14 (b) | $t\bar{t}$ | 0.900±0.100 | 1.290±0.300 | 0.875±0.015 | 0.540±0.010 | 0.070 |
| 15 (c) | $l$ | 0.905±0.100 | 1.280±0.260 | 1.140±0.025 | 0.570±0.018 | 0.232 |
| 16 (c) | $b$ | 0.923±0.112 | 1.280±0.255 | 1.240±0.030 | 0.550±0.015 | 0.027 |

Table 3: The values of $\langle p_T \rangle$ and $\sqrt{\langle p_T^2 \rangle / 2}$ are calculated by the curves in Figure 1, 3, 5, 7.

| Figure | Type | $\langle p_T \rangle$ (GeV/c) | $\sqrt{\langle p_T^2 \rangle / 2}$ (GeV/c) |
|---|---|---|---|
| | at parton level in the full phase space | | |
| 9(a)/10(a) | $t$ | 120.691±0.014 | 102.471±6.971 |
| | $\bar{t}$ | 119.785±0.012 | 101.604±6.247 |
| | $t$ (leading) | 137.210±0.015 | 114.073±7.430 |
| | $t$ (trailing) | 103.660±0.006 | 88.028±3.131 |
| | $t\bar{t}$ | 112.256±0.008 | 94.944±4.114 |
| | at particle level in the fiducial phase space | | |
| 9(b)/10(b) | $t$ | 126.834±0.017 | 108.302±8.416 |
| | $\bar{t}$ | 127.002±0.018 | 108.153±8.762 |
| | $t$ (leading) | 136.458±0.022 | 115.996±10.800 |
| | $t$ (trailing) | 108.473±0.008 | 92.235±3.834 |
| | $t\bar{t}$ | 119.830±0.012 | 101.536±5.958 |
| 9(c)/10(c) | $l$ | 53.617±0.004 | 49.657±1.768 |
| | $\bar{l}$ | 53.322±0.003 | 49.075±1.703 |
| | $l$ (leading) | 64.970±0.005 | 58.171±2.743 |
| | $l$ (trailing) | 34.914±0.001 | 32.255±0.450 |
| | $l\bar{l}$ | 81.272±0.003 | 68.092±1.428 |
| 9(d)/10(d) | $b$ (leading) | 69.613±0.004 | 62.890±1.861 |
| | $b$ (trailing) | 42.804±0.001 | 39.883±0.348 |
| | $b\bar{b}$ | 69.127±0.002 | 61.179±0.867 |

Table 4: Values of intercepts, slopes and $\chi^2/dof$ corresponding to the lines in Figure 17.

| Particle | Intercept | Slope | $\chi^2/dof$ |
|---|---|---|---|
| | $T - \sqrt{s}$ | | |
| t | 21.068±1.276 | 0.638±0.132 | 0.113 |
| b | 26.846±3.547 | 0.066±0.366 | 0.796 |
| l | 6.835±1.149 | 0.135±0.118 | 1.643 |
| | $n - \sqrt{s}$ | | |
| t | 9.380±0.588 | -0.015±0.061 | 0.258 |
| b | 25.702±3.888 | -0.848±0.401 | 3.324 |
| l | 3.886±1.025 | 0.071±0.106 | 1.996 |
| | $\sigma(C) - \sqrt{s}$ | | |
| t | 0.757±0.111 | 0.037±0.011 | 9.802 |
| b | 1.210±0.054 | 0.000±0.006 | 1.405 |
| l | 0.987±0.001 | 0.019±0.000 | 0.001 |
| | $\sigma(P/T) - \sqrt{s}$ | | |
| t | 0.283±0.232 | 0.035±0.024 | 158.376 |
| b | 0.626±0.163 | 0.001±0.017 | 41.913 |
| l | 0.644±0.140 | 0.001±0.014 | 22.677 |

**4. Conclusion**

We summarize here our main observations and conclusions.

(a) We use the Tsallis-Pareto-type function to describe the transverse-momentum cross-sections spectrum of $t\bar{t}$ different cross section in *pp* collisions at 7, 8 and 13 TeV. The values of the effective temperature of the interacting system (T) and the non-extensivity of the process (n) parameters are extracted and listed in Table 1.

(b) The rapidity cross-sections of the spectrum of $t\bar{t}$ different cross section in *pp* collisions at 7, 8 and 13 TeV are analyzed by the three-source Landau hydrodynamic model. We have obtained the values of the contribution of central emission source ($k_C$), the width of rapidity distribution ($\sigma_C, \sigma_P$) and calculated the values of $\chi^2/dof$. The values of related parameters and $\chi^2/dof$ are given in Table 2. In fact, because of the relationship $k_T + k_C + k_P = 1$, $k_T = k_P$ and $\sigma_T = \sigma_P$ (for the symmetric collision), we can calculate the contributions of target and projectile source and the width of rapidity distribution of target source.

(c) We have ploted the relationship of free parameters and collision energy in Figure 17. For the particles of *t*, *l* and *b*, the $\sigma(C)$ shows an increasing trend and T shows a slight increased with the collision energy increased.

(d) As has been mentioned in section 3, the initial temperature ($T_i$) can be described by $\sqrt{\langle p_T^2 \rangle/2}$ approximately. In order to understand the excitation degree of the interacting system, we calculated the values of $\langle p_T \rangle$ and $\sqrt{\langle p_T^2 \rangle/2}$ by the curves in Figure 1, 3, 5, 7. We found the values of $\langle p_T \rangle$ and $\sqrt{\langle p_T^2 \rangle/2}$ are large, this could mean the high excitation degree of the interacting system.

**Data Availability**
The data used to support the findings of this study are included within the article.

**Ethical Approval**
The authors declare that they are incompliance with ethical standards regarding the content of this paper.

**Conflicts of Interest**
The authors declare that they have no conflicts of interest regarding the publication of this paper.

**Acknowledgments**
This work is supported by Scientific and Technological Innovation Programs of Higher Education Institutions in Shanxi under Grant No. 2019L0804, the National Natural Science Foundation of China under Grant No.11847003, and the university student innovation and entrepreneurship training program of Taiyuan Normal University No. CXCY 2084.